\documentclass[aps,pra,amsmath,amssymb,twocolumn,superscriptaddress,showpacs]{revtex4-1}
\usepackage{ulem}
\usepackage{amsmath,amssymb}
\usepackage{graphicx}	

\mathchardef\mhyphen="2D

\def\be{\begin{eqnarray}}   
\def\ee{\end{eqnarray}}
\def\vecb{\boldsymbol}
\def\MIR{\mathrm{MIR}}
\def\THz{\mathrm{THz}}
\def\dc{\mathrm{dc}}
\def\eff{\mathrm{eff}}

\begin{document}

\author{Wenwen~Mao}
\affiliation 
{Max Planck Institute for the Structure and Dynamics of Matter, Luruper Chaussee 149, 22761 Hamburg, Germany}

\author{Angel~Rubio}
\affiliation 
{Max Planck Institute for the Structure and Dynamics of Matter, Luruper Chaussee 149, 22761 Hamburg, Germany}
\affiliation 
{Center for Computational Quantum Physics, Flatiron Institute, 162 Fifth Avenue, New York, NY 10010, USA}

\author{Shunsuke~A.~Sato}
\email{ssato@ccs.tsukuba.ac.jp}
\affiliation 
{Center for Computational Sciences, University of Tsukuba, Tsukuba 305-8577, Japan}
\affiliation 
{Max Planck Institute for the Structure and Dynamics of Matter, Luruper Chaussee 149, 22761 Hamburg, Germany}

\title{Enhancement of high-order harmonic generation in graphene by mid-infrared and terahertz fields}

\begin{abstract}
We theoretically investigate high-order harmonic generation (HHG) in graphene under mid-infrared (MIR) and terahertz (THz) fields based on a quantum master equation. Numerical simulations show that MIR-induced HHG in graphene can be enhanced by a factor of 10 for fifth harmonic and a factor of 25 for seventh harmonic under a THz field with a peak strength of 0.5~MV/cm by optimizing the relative angle between the MIR and THz fields. To identify the origin of this enhancement, we compare the fully dynamical calculations with a simple thermodynamic model and a nonequilibrium population model. The analysis shows that the enhancement of the high-order harmonics mainly results from a coherent coupling between MIR- and THz-induced transitions that goes beyond a simple THz-induced population contribution.
\end{abstract}

\maketitle

\section{Introduction \label{sec:intro}}

Recent developments in laser technologies have enabled the generation of intense light~\cite{RevModPhys.72.545,RevModPhys.81.163,MOUROU2012720}, opening an avenue into the investigation of light-matter interactions in nonlinear regimes. High-order harmonic generation (HHG) is an extremely nonlinear optical phenomenon involving extreme photon upconversion. This phenomenon was first observed in atomic gases~\cite{McPherson:87,Ferray_1988}. A semi-classical three-step model has provided a clear understanding of gas-phase HHG~\cite{PhysRevLett.71.1994,PhysRevA.49.2117}. Gas-phase HHG has been utilized to generate ultrashort attosecond light pulses~\cite{RevModPhys.81.163} for investigating light-induced ultrafast electron dynamics in atoms~\cite{Goulielmakis2010,PhysRevLett.105.143002,PhysRevLett.106.123601}, molecules~\cite{Warrick2016,Reduzzi2016,PhysRevResearch.3.043222}, and solids~\cite{doi:10.1126/science.1260311,doi:10.1126/science.aag1268,Mashiko2016,Siegrist2019}. HHG from a solid-state system was first observed in 2011~\cite{Ghimire2011} and has recently been attracting considerable interest from both fundamental and technological points of view~\cite{Ghimire2019,Silva2019,Nakagawa2022,gorlach2022high,neufeld2023universal}.

Among various materials, the HHG from graphene has been intensively investigated theoretically. It has been suggested that HHG can be efficiently induced in graphene because of the unique electronic structure of this material, that is, the Dirac cone~\cite{Mikhailov_2007,PhysRevB.82.201402,PhysRevB.90.245423,PhysRevB.89.041408}. Recently, HHG from graphene has been experimentally observed in the mid-infrared (MIR)~\cite{doi:10.1126/science.aam8861,cha2022gate} and terahertz (THz)~\cite{Hafez2018,doi:10.1126/sciadv.abf9809} regimes, exhibiting unique ellipticity dependence and high efficiency. We previously investigated HHG from graphene in both the MIR and THz regimes based on a quantum master equation~\cite{PhysRevB.103.L041408,PhysRevB.106.024313}. In the MIR regime, coupling between field-induced intraband and interband transitions opens important channels for HHG, enhancing HHG with finite ellipticity~\cite{PhysRevB.103.L041408}. A real-time electron dynamics simulation in the THz regime has shown that it is essential to consider the nonequilibrium steady-state resulting from the balance between field-driving and relaxation to go beyond the equilibrium thermodynamic picture of HHG from graphene~\cite{PhysRevB.106.024313}.

It is important to improve the efficiency of solid-state HHG to develop novel HHG-based light sources and spectroscopies. In recent studies, it has been suggested that HHG from graphene can be enhanced using two-color laser fields based on various mechanisms~\cite{PhysRevB.100.035434,Mrudul:21,PhysRevB.105.195405}. In Ref.~\cite{PhysRevB.100.035434}, the two-color HHG is suggested with the combination of the electron-hole pair creation by high-frequency pump light and the acceleration of the created pairs by low-frequency light. Mrudul \textit{et al} investigated the HHG from graphene with bicircular fields, controlling the valley polarization~\cite{Mrudul:21}. Furthermore, Avetissian~\textit{et al} investigated the HHG from graphene with a linearly polarized light and its second harmonics, showing that when the two-color fields are perpendicularly polarized, the stronger harmonics can be emitted compared with the parallel polarization~\cite{PhysRevB.105.195405}.

In this study, we explore the possibility of using a THz field to enhance MIR-induced HHG in graphene based on the knowledge gained from previous studies. First we use a quantum master equation to compute the electron dynamics under MIR and THz fields and evaluate the emitted harmonic spectra. We compare the results of the fully dynamical calculations with a thermodynamic model considering the equilibrium Fermi--Dirac distribution and with a nonequilibrium population model considering a population distribution in a nonequilibrium steady-state. We find that coherent driving by the THz field plays an essential role in enhancing MIR-induced HHG beyond the induced population effect.

The paper is organized as follows. In Sec.~\ref{sec:method}, we briefly describe the theoretical method for describing electron dynamics in graphene induced by MIR and THz fields. In Sec.~\ref{sec:results}, we investigate the impact of a THz field on MIR-induced HHG from graphene and explore the microscopic mechanism of the enhancement by employing the thermodynamic model and the nonequilibrium population model. Finally, our findings are summarized in Sec.~\ref{sec:summary}.

\section{Methods \label{sec:method}}

\subsection{Electron dynamics calculation based on a quantum master equation \label{subsec:qmaster}}

Here, we briefly describe theoretical methods for computing light-induced electron dynamics in graphene and how the calculated dynamics can be used to analyze HHG. These methods have been described in detail in previous works~\cite{PhysRevB.99.214302,sato2019light,PhysRevB.106.024313}.

In this study, we use the following quantum master equation to describe electron dynamics in graphene:
\begin{align}
\frac{\mathrm{d}}{\mathrm{d}t}\rho_{\vecb{k}}(t) = \frac{1}{i \hbar}	\left[ H_{\vecb{k}+e\vecb{A}(t)/\hbar}, \rho_{\vecb{k}} (t)) \right] + 	
\hat{D}\left[ \rho_{\vecb{k}} (t)) \right],
\label{eqn:masterequation}
\end{align}
where $\rho_{\vecb k}(t)$ is the one-body reduced density matrix for the Bloch wavevector $\vecb k$, and $H_{\vecb k+e\vecb{A}(t)/\hbar}$ is the one-body Hamiltonian. Here, light-matter coupling is described by an additional term in the Hamiltonian, a spatially uniform vector potential $\vecb A(t)$, which is related to the applied electric field as $\vecb A(t)=-\int^t_{-\infty}dt' \vecb E(t')$, in the long-wavelength approximation. The relaxation operator is denoted as $\hat{D}\left[ \rho_{\vecb{k}} (t)) \right]$. Based on the previous studies~\cite{PhysRevB.99.214302, PhysRevB.106.024313}, we employ the relaxation time approximation~\cite{PhysRevLett.73.902} for the relaxation operator in the Houston basis expression~\cite{PhysRev.57.184,PhysRevB.33.5494}. Those Houston states $|u^H_{b\vecb k}(t)\rangle$ are simply the instantaneous eigenstates of the time-dependent Hamiltonian and satisfy the following instantaneous eigenvalue problem:
\begin{align}
H_{\vecb{k}+e\vecb{A}(t)/\hbar} |u^H_{b\vecb k}(t)\rangle = \epsilon_{b,\vecb k + e\vecb A(t)/\hbar}|u^H_{b\vecb k}(t)\rangle,
\end{align}
where $b$ denotes the band index, and $\epsilon_{b,\vecb k + e\vecb A(t)/\hbar}$ are the instantaneous eigenvalues of the Hamiltonian. The reduced density matrix can be expanded in the Houston states as
\begin{align}
  \rho_{\vecb k}(t) = \sum_{bb'}\rho_{bb',\vecb k}(t)|u^H_{b\vecb k}(t)\rangle \langle u^H_{b'\vecb k}(t)|,
\end{align}
where $\rho_{bb',\vecb k}(t)$ are the expansion coefficients. The relaxation operator $\hat{D}\left[\rho_{\vecb{k}} (t)\right]$ can also be expanded in the Houston states as
\begin{widetext}
\begin{align}
  \hat D\left [\rho_{\vecb k}(t) \right ]=-\sum_{b}\frac{\rho_{bb,\vecb k}(t)-f^{FD}\left (\epsilon_{b,\vecb k+e\vecb A(t)/\hbar},T_e,\mu \right)}{T_1}|u^H_{b\vecb k}(t)\rangle \langle u^H_{b\vecb k}(t)| 
  -\sum_{b\neq b'} \frac{\rho_{bb',\vecb k}(t)}{T_2}|u^H_{b\vecb k}(t)\rangle \langle u^H_{b'\vecb k}(t)|.
\label{eqn:relaxation}
\end{align}
\end{widetext}

From Refs.~\cite{PhysRevB.99.214302, PhysRevB.106.024313}, the longitudinal relaxation time $T_1$ is set to 100~fs, and the transverse relaxation time $T_2$ is set to 20~fs. Here, $f^{\mathrm{FD}}(\epsilon, T_e, \mu)=\frac{1}{e^{(\epsilon-\mu)/k_BT_e}+1}$ is the Fermi--Dirac distribution, in which we set the electron temperature $T_e$ to 300~K and the chemical potential $\mu$ to 0, unless stated otherwise.

We describe the electronic structure of graphene employing a tight-binding Hamiltonian with nearest-neighbor hopping~\cite{RevModPhys.81.109} as follows:
\begin{align}
H_{\vecb{k}}=\left(\begin{array}{cc}
0 & t_{0} f(\vecb{k}) \\
t_{0} f(\vecb{k})^{*} & 0
\end{array}\right),
\label{eqn:TBhamiltonian}
\end{align}
where $t_0$ is the nearest-neighbor hopping, and $f(\boldsymbol{k})$ is given by $f(\boldsymbol{k})=e^{i \boldsymbol{k} \cdot \vecb{\delta}_{1}}+e^{i \boldsymbol{k} \cdot \vecb{\delta}_{2}}+e^{i \boldsymbol{k} \cdot \vecb{\delta}_{3}}$ with the nearest-neighbor vector $\vecb \delta_j$. According to Ref.~[\cite{RevModPhys.81.109}], the nearest-neighbor hopping $t_0$ is set to 2.8~eV, and the lattice constant $a$ is set to $1.42$~$\AA$.

By using the time-dependent density matrix $\rho_{\vecb k}(t)$ evolved with Eq.~(\ref{eqn:masterequation}), the total energy of the electronic system can be evaluated as
\begin{eqnarray}
E_{\mathrm{tot}}(t)=\frac{2}{\Omega_{\mathrm{BZ}}} \int d\vecb k \mathrm{Tr}\left[H_{\vecb k+ e\vecb A(t)/\hbar} \rho_{\vecb k}(t)\right],
\label{eqn:totalenergy}
\end{eqnarray}
where $\Omega_{\mathrm{BZ}}$ is the volume of the first Brillouin zone.

Similarly, the electric current density is given by
\begin{eqnarray}
  \vecb{J}(t)=\frac{2}{(2\pi)^2} \int d\vecb k \mathrm{Tr}\left[\hat{\boldsymbol{J}}_{\boldsymbol{k}}(t)\rho_{\boldsymbol{k}}(t)\right],
\label{eqn:totalcurrent}
\end{eqnarray}
where the current operator $\hat{\vecb J}_{\vecb k}(t)$ is defined as
\begin{align}
\hat{\vecb{J}}_{\vecb{k}}(t) =  -\frac{e}{m_e \hbar}\frac{\partial H_{\vecb k + e\vecb A(t)/\hbar}}{\partial \vecb k}.
\end{align}

The current density $\vecb{J}(t)$ induced by an intense electric field $\vecb E(t)$ is analyzed to investigate HHG. For example, the power spectrum of the emitted harmonics can be evaluated by applying the Fourier transform to the current density $\vecb J(t)$ as follows:
\begin{align}
I_{\mathrm{HHG}}(\omega)\sim \omega^2 \left | \int^{\infty}_{-\infty} dt \vecb J(t) e^{i\omega t} \right |^2.
\label{eqn:spectrum}
\end{align}

\section{Results \label{sec:results}}

\subsection{MIR-induced HHG in graphene under THz fields and the quasistatic approximation \label{subsec:quasistatic}}

We first analyze the HHG induced by a MIR laser pulse in the presence of THz fields. For practical calculations, we employ the following form for the MIR pulse:
\begin{align}
  \vecb A_{\MIR}(t) = -\frac{E_{\MIR}}{\omega_{\MIR}} \vecb{e}_{\MIR} 
\sin(\omega_{\MIR} t) \cos^4 \left (\frac{\pi}{T_{\MIR}} t \right)
  \label{eqn:laser_pulse}
\end{align}
in the domain $-T_{\MIR}/2<t<T_{\MIR}/2$, and zero outside this domain. Here, $E_{\MIR}$ is the peak strength of the MIR field, $\omega_{\MIR}$ is the mean frequency, $\vecb e_{\MIR}$ is a unit vector along the polarization direction of light, and $T_{\MIR}$ is the pulse duration. In this study, the pulse duration $T_{\MIR}$ is set to 40~ps, and the mean frequency $\omega_{\MIR}$ is set to 0.35424~eV/$\hbar$. We compute the electron dynamics by changing the other parameters.

\begin{figure*}[ht]
\includegraphics[width=0.9\linewidth]{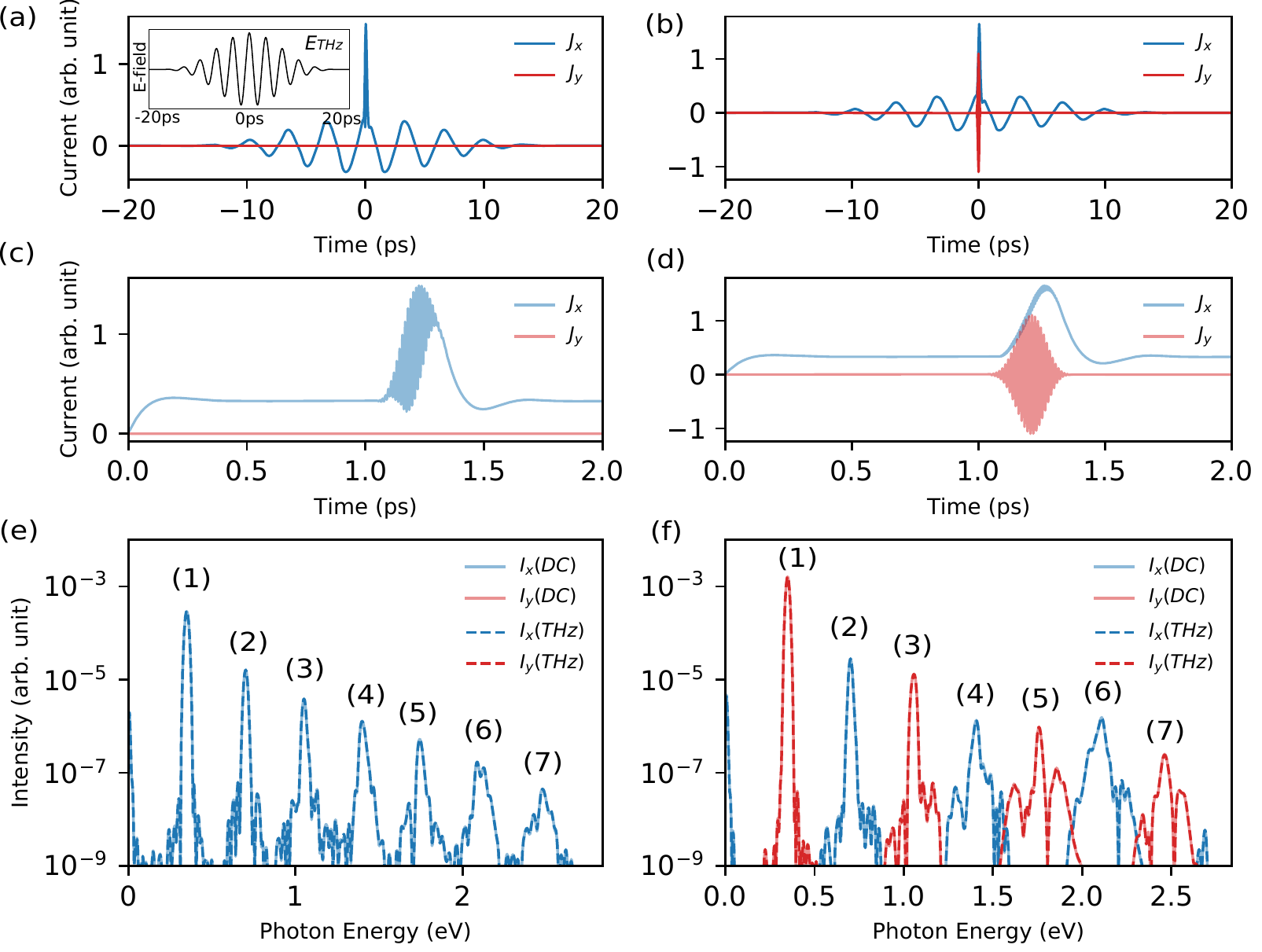}
\caption{\label{fig:current}
(a, b)~The current $\vecb{J} (t)$ induced by THz and MIR fields, $\vecb E_{\THz}(t)+\vecb E_{\MIR}(t)$. The inset is the panel (a) shows the time profile of the applied THz field. (c, d)~The current $\vecb{J}(t)$ induced by the static and MIR fields, $\vecb E_{\dc}(t)+\vecb E_{\MIR}(t-\tau_{\MIR})$. In the panels~(a) and (c), the polarization of all the fields is parallel to the $\Gamma$--$M$ direction (the $x$-direction in the present setup) as $\vecb e_{\THz}=\vecb e_{\dc}=\vecb e_{\MIR}=\vecb e_x$. In the panels~(b) and (d), the polarization of THz and static fields is parallel to the $x$-direction as $\vecb e_{\THz}=\vecb e_{\dc}=\vecb e_{\MIR}=\vecb e_x$, while that of the MIR field is perpendicular as $\vecb e_{\MIR}=\vecb e_y$. (e) The power spectra $I_{\mathrm{HHG}}(\omega)$ computed using the current in (a) and (c). (f) The power spectra $I_{\mathrm{HHG}}(\omega)$ computed using the current in (b) and (d).
}
\end{figure*}

Similarly, we employ the following form for the THz pulse,
\begin{align}
  \vecb A_{\THz}(t) = -\frac{E_{\THz}}{\omega_{\THz}} \vecb{e}_{\THz} 
\sin(\omega_{\THz} t) \cos^4 \left (\frac{\pi}{T_{\THz}} t \right)
  \label{eqn:laser_pulse}
\end{align}
in the domain $-T_{\THz}/2<t<T_{\THz}/2$, and zero outside this domain. Here, $E_{\THz}$ is the peak strength of the THz field, $\omega_{\THz}$ is the mean frequency, $\vecb e_{\THz}$ is a unit vector along the polarization direction, and $T_{\THz}$ is the pulse duration. In this study, the pulse duration $T_{\THz}$ is set to 42~ps and the mean frequency $\omega_{\THz}$ is set to 1.2407~meV/$\hbar$. The time profile of the applied THz electric field is shown in the inset of Fig.~\ref{fig:current}~(a). 

To gain insight into THz-assisted MIR-induced HHG in graphene, we perform the electron dynamics calculation in the presence of both THz and MIR fields, $\vecb E_{\THz}(t)+\vecb E_{\MIR}(t)$. Here, we set $E_{\MIR}$ to 6.5~MV/cm and $E_{\THz}$ to 0.5~MV/cm. We note that intense THz pulses with amplitudes exceeding 1~MV/cm are available experimentally~\cite{10.1063/1.3560062}. The polarization direction of the THz field $\vecb e_{\THz}$ is set to the $\Gamma$--$M$ direction (the $x$-direction in our setup), whereas the polarization direction of the MIR field $\vecb e_{\MIR}$ is treated as a tunable parameter. Figures~\ref{fig:current}~(a) and (b) show the computed current $\vecb J(t)$ induced by $\vecb E_{\THz}(t)+\vecb E_{\MIR}(t)$ as a function of time. The result with the parallel configuration ($\vecb e_{\MIR}=\vecb e_x = \vecb e_{\THz}$) is shown in Fig.~\ref{fig:current}~(a), while the result with the perpendicular configuration ($\vecb e_{\MIR}=\vecb e_y \perp \vecb e_{\THz}$) is shown in Fig.~\ref{fig:current}~(b). As seen from Figs.~\ref{fig:current}~(a) and (b), the THz field induces a current on the picosecond time scale, whereas the MIR field induces a current on a much shorter time scale.

We analyze MIR-induced HHG by extracting the current induced by the MIR field in the presence of the THz field. We define two kinds of currents for this purpose. The first current is induced by both the THz and MIR fields and is denoted as $\vecb J^{\THz + \MIR}(t)$. The second current is induced solely by the THz field and is denoted as $\vecb J^{\THz}(t)$. We define the current induced by the MIR field in the presence of the THz field as $\vecb J^{\eff}(t)=\vecb J^{\THz + \MIR}(t)-\vecb J^{\THz}(t)$. The Fourier transform is applied to the extracted current $\vecb J^{\eff}(t)$, and the power spectrum of the emitted harmonics is computed using Eq.~(\ref{eqn:spectrum}). The dashed line in Fig.~\ref{fig:current}~(e) shows the power spectrum computed using the current $\vecb J(t)$ presented in Fig.~\ref{fig:current}~(a), where the polarization directions of the THz and MIR fields are parallel, and the dashed line in Fig.~\ref{fig:current}~(f) shows the power spectrum computed using the current $\vecb J(t)$ presented in Fig.~\ref{fig:current}~(b), where the polarization directions of the two fields are perpendicular. Figure~\ref{fig:current}~(e) shows that second and higher even-order harmonics are generated in addition to odd-order harmonics because the THz field breaks the inversion symmetry of the system locally in time. This second-harmonic generation is known as electric-field-induced second-harmonic generation (EFISH) or THz-induced second-harmonic generation (TFISH)~\cite{PhysRevLett.8.404,PhysRev.137.A801,Nahata:98,COOK1999221}. Even-order harmonics are similarly generated in the perpendicular configuration ($\vecb e_{\MIR} \perp \vecb e_{\THz}$), as can be seen from Fig.~\ref{fig:current}~(f).

Explicit use of the THz pulse in the electron dynamics calculation increases the propagation time (4000~fs in the present case), as can be seen Figs.~\ref{fig:current}~(a) and (b). Hence, the electron dynamics calculation with the explicit inclusion of THz pulses has a large computational cost. To lower the computational cost of modeling MIR-induced HHG in graphene in the presence of a THz field, we replace the THz pulse with a static field to calculate the electron dynamics, corresponding to a quasistatic approximation~\cite{PhysRevB.106.024313}. For practical analysis, we perform two simulations. In the first simulation, the electron dynamics are calculated under a static field $\vecb E_{\dc}(t)=\vecb e_{\dc} E_{\dc}\Theta(t)$ that is suddenly switched on at $t=0$. Here, $\vecb e_{\dc}$ is the unit vector along the polarization direction of the static field, and $E_{\dc}$ is the field strength. Immediately after the static field is switched on, the driven electron dynamics induce a current. The driven system reaches a steady state after a sufficiently long time propagation time, and the current becomes constant in time. We denote the current induced solely by $\vecb E_{\dc}(t)$ as $\vecb J^{\dc}(t)$. In the second simulation, the electron dynamics are calculated under both MIR and static fields, $\vecb E_{\dc}(t)+\vecb E_{\MIR}(t-\tau_{\MIR})$, where the pulse center of the MIR field is shifted by $\tau_{\MIR}$. We denote the current induced by $\vecb E_{\dc}(t)+\vecb E_{\MIR}(t-\tau_{\MIR})$ as $\vecb J^{\dc+\MIR}(t)$. The shift $\tau_{\MIR}$ can be made sufficiently large time to investigate the MIR-induced electron dynamics for a full nonequilibrium steady state realized by the static field $\vecb E_{\dc}(t)$. The MIR-induced current can be extracted as $\vecb J^{\eff}(t)=\vecb J^{\dc+\MIR}(t)-\vecb J^{\dc}(t)$ to analyze MIR-induced HHG in the presence of the static field.

Figures~\ref{fig:current}~(c) and (d) show the current $\vecb J^{\dc+\MIR}(t)$ induced by both static and MIR fields. Here, the static field is polarized along the $\Gamma$--$M$ direction (the $x$-direction in our setup), and the field strength $E_{\dc}$ is the same as the peak strength of the THz field, $E_{\dc}=E_{\THz}=0.5$~MV/cm. The MIR field used in Fig.~\ref{fig:current}~(c) is the same as that used in Fig.~\ref{fig:current}~(a) and has a polarization direction parallel to that of the static field. By contrast, the MIR field used in Fig.~\ref{fig:current}~(d) is the same as that used in Fig.~\ref{fig:current}~(b) and has a polarization direction perpendicular to that of the static field. To apply the MIR field to the nonequilibrium steady-state under the static field, we set the time delay $\tau_{\MIR}$ of the MIR field to 1~ps, which is sufficiently longer than the relaxation time scales of the quantum master equation, $T_1$ and $T_2$.

To analyze HHG in the presence of the static field $\vecb E_{\dc}(t)$, we extract the current $\vecb J^{\eff}(t)$ induced by the MIR field in the presence of the static field by subtracting $\vecb J^{\dc}(t)$ from $\vecb J^{\dc+\MIR}(t)$: $\vecb J^{\eff}(t)=\vecb J^{\dc+\MIR}(t)-\vecb J^{\dc}(t)$. The solid lines in Figs~\ref{fig:current}~(e) and (f) correspond to the HHG spectra computed using the current shown in Figs~\ref{fig:current}~(c) and (d), respectively. The results of the quasistatic approximation are identical to those computed by explicitly including the THz pulse. Therefore, the quasistatic approximation is very good for analyzing HHG under MIR and THz fields. Hereafter, we employ the static field within the quasistatic approximation instead of explicitly including the THz pulse. The agreement between results obtained using the quasistatic approximation and the explicit inclusion of the THz pulse indicates that the nonequilibrium steady state under the static field plays an important role in MIR-induced HHG in graphene in the presence of a THz field.

\subsection{Orientational dependence of HHG}

Here, we investigate HHG in graphene within the quasistatic approximation by changing the relative angle between the static and MIR fields. For practical analysis, the direction of the static field $\vecb e_{\dc}$ is fixed to the $\Gamma$--$M$ axis (the $x-$axis in our setup), and the peak field strength of the MIR field $E_{\MIR}$ is fixed at 6.5~MV/cm. The emitted harmonics are investigated by manipulating the polarization direction of the MIR field, $\vecb e_{\MIR}$, and the strength of the static field, $E_{\dc}$.

To analyze the HHG efficiency, we compute the signal intensity of the emitted harmonics at each order by integrating the power spectrum within a finite energy range as follows:
\begin{align}
I^{n \textrm{th}}_{\mathrm{total}} = \int_{\left (n-\frac{1}{2} \right )\omega_{\MIR}}^{\left (n+\frac{1}{2} \right )\omega_{\MIR}} d \omega I_{\textrm{HHG}} (\omega).
\label{eqn:integrate_intensity}
\end{align}
Here, $I^{n \textrm{th}}_{\mathrm{total}}$ is the integrated intensity of the emitted $n$th harmonic.

\begin{figure*}[ht]
\centering
\includegraphics[width=0.95\linewidth]{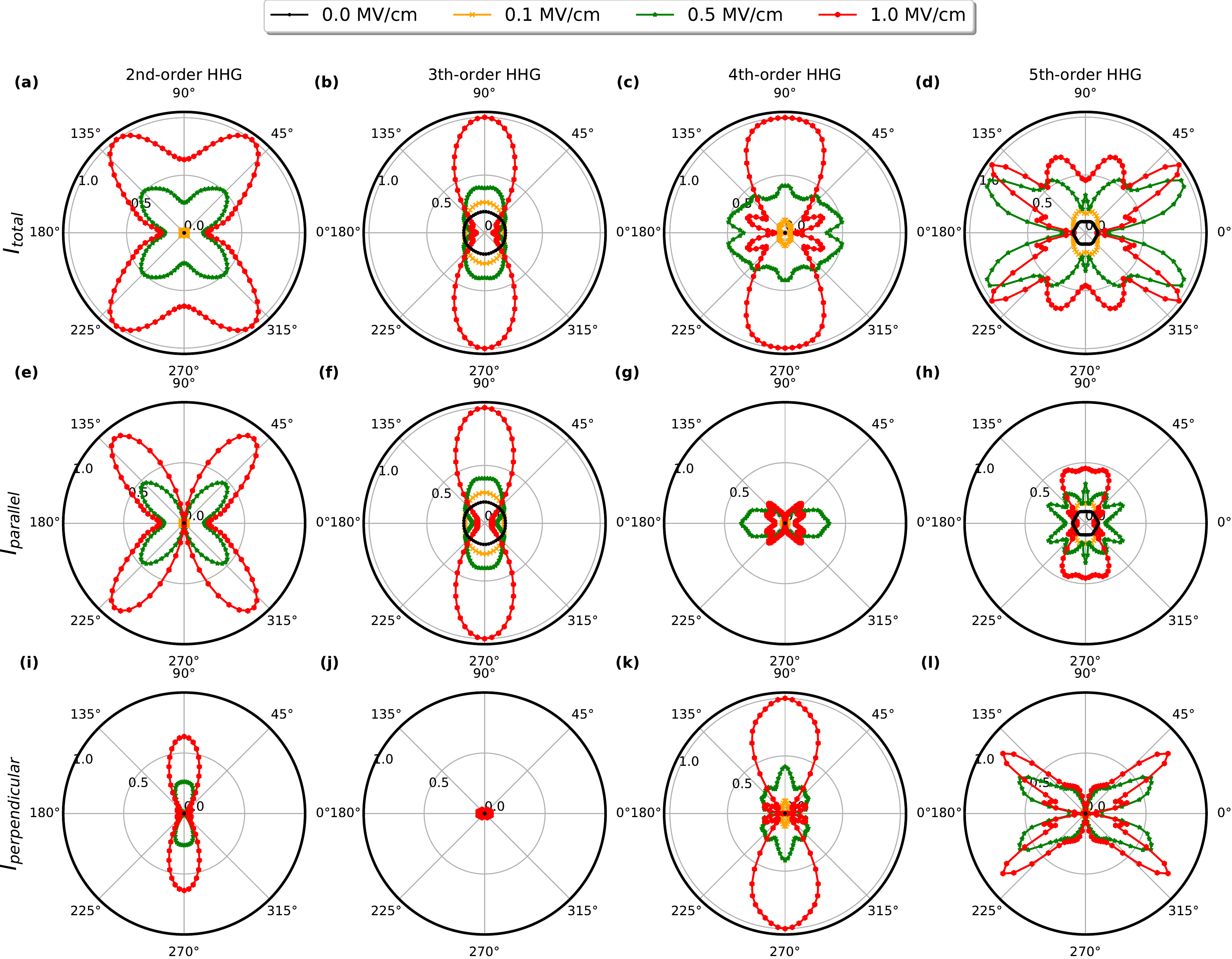}
\caption{\label{fig:polar}
The angular dependence of the harmonic yield in the nonequilibrium steady-states under a static field along the $\Gamma$--$M$ direction is shown. The angle $\theta$ denotes the relative angle between the static field and the $\MIR$ field. (a--d) The total intensity $I^{n \textrm{th}}_{\mathrm{total}}$ is shown for the second, third, fourth, and fifth harmonics. (e-h) The component of the intensity parallel to $\vecb e_{\MIR}$ is shown for each harmonic. (i-l) The component of the intensity perpendicular to $\vecb e_{\MIR}$is shown for each harmonic. The results are normalized by the maximum total intensity $I^{n \textrm{th}}_{\mathrm{total}}$ for each harmonic.
}
\end{figure*}

Figures~\ref{fig:polar}~(a--d) show the computed angular dependence of the emitted harmonic yield $I^{n \textrm{th}}$ for different harmonic orders. The angle $\theta$ denotes the relative angle between the MIR and static fields. In Fig.~\ref{fig:polar}~(a), there is no second harmonic because graphene has intrinsic inversion symmetry in the absence of a static field. By contrast, the second harmonics are generated under the application of a static field because of the breakdown of the inversion symmetry. For a static field strength of 0.5~MV/cm, the emitted second-harmonic intensity is maximized at a relative angle of approximately 45$^{\circ}$.

In Fig.~\ref{fig:polar}~(b), the third-harmonic yield is almost isotropic (black line) in the absence of a static field, reflecting the rotational symmetry of the Dirac cone (see also Appendix~\ref{sec:angle-dep-higher-wo-thz}). By contrast, the third-harmonic intensity exhibits a strong angular dependence under the application of a strong static field ($E_{\dc}=1.0$~MV/cm): the third-harmonic emission is considerably enhanced when the static and MIR fields are perpendicular to each other and suppressed when these two fields are parallel. The enhancement of the third harmonic for the perpendicular configuration can be understood in terms of the coupling between the intraband transition induced by the static field and the interband transition induced by the MIR field, as was suggested in a previous study~\cite{PhysRevB.103.L041408}.

By comparison, the higher-order harmonics exhibit a more complex angular dependence under a static field, as shown in Fig.~\ref{fig:polar}~(c and d). In particular, the fifth-order harmonic emission can be considerably enhanced in the presence of static or THz fields (Fig.~\ref{fig:polar}~(d)). For example, the intensity of the fifth-order harmonic is enhanced more than ten times by applying a static field with a strength of 0.5~MV/cm with respect to the result solely induced by the MIR field (see the green line in Fig.~\ref{fig:polar}~(d)). This enhancement ratio is larger than that of the third-order harmonic. Hence, a larger field-induced enhancement is expected for higher-order harmonics. In fact, the seventh-order harmonic is enhanced 25 times when the static field strength is 0.5~MV/cm (see Appendix~\ref{sec:angle-dep-higher}).

To further elucidate the angular dependence of HHG in graphene, we decompose the harmonic intensity $I_{\mathrm{HHG}}(\omega)$ into parallel and perpendicular components with respect to the polarization of the driving MIR field. The parallel component of the HHG intensity is defined as 
\begin{align}
I^{\textrm{para}}_{\mathrm{HHG}}(\omega)\sim \omega^2 \left | \int^{\infty}_{-\infty} dt \vecb e_{\MIR} \cdot \vecb J(t) e^{i\omega t} \right |^2,
\label{eqn:spectrum-para}
\end{align}
where $\vecb e_{\MIR}$ is the unit vector along the polarization direction of the MIR field. Likewise, the perpendicular component is defined as
\begin{align}
I^{\textrm{perp}}_{\mathrm{HHG}}(\omega)\sim \omega^2 \left | \int^{\infty}_{-\infty} dt \bar{\vecb e}_{\MIR} \cdot \vecb J(t) e^{i\omega t} \right |^2,
\label{eqn:spectrum-perp}
\end{align}
where $\bar{\vecb e}_{\MIR}$ is a unit vector perpendicular to $\vecb e_{\MIR}$, i.e., $\bar{\vecb e}_{\MIR} \cdot\vecb e_{\MIR}=0$. The total intensity $I_{\textrm{HHG}}$ in Eq.~(\ref{eqn:spectrum}) is reproduced by the sum of $I^{\textrm{para}}_{\mathrm{HHG}}(\omega)$ and $I^{\textrm{perp}}_{\mathrm{HHG}}(\omega)$ as $I_{\mathrm{HHG}}(\omega)=I^{\textrm{para}}_{\mathrm{HHG}}(\omega)+I^{\textrm{perp}}_{\mathrm{HHG}}(\omega)$.

Eq.~(\ref{eqn:spectrum-para}) and Eq.~(\ref{eqn:spectrum-perp}) are directionally decomposed to analyze the parallel and perpendicular components of the emitted harmonics with respect to the polarization direction of the MIR field. Figures~\ref{fig:polar}~(e--h) and~(i--l) show the angular dependence of the parallel and perpendicular components the harmonic intensity for orders, respectively.

In Figs~\ref{fig:polar}~(a), (e), and (i), the parallel component of the second harmonic under the static field reaches a maximum at approximately 45$^{\circ}$ and dominates the total second-harmonic intensity at this angle. By contrast, the maximum perpendicular component is always obtained when the MIR and static fields are perpendicular to each other. In Figs.~\ref{fig:polar}~(b), (f), and (j), the third harmonic is dominated by the parallel component for any angle and static field strength over the investigated range. For both second- and third-harmonic generation, the parallel components are dominant when the emitted harmonic intensity is maximized.

There are qualitative differences between the lower-order harmonics (the second- and third-order harmonics) and the higher-order harmonics (the fourth- and fifth-order harmonics). In Fig.~\ref{fig:polar}~(c), the fourth harmonic yield reaches a maximum at an angle $\theta$ of 90$^{\circ}$ under the strongest applied static field, $E_{\dc}=1.0$~MV/cm. A comparison of Fig.~\ref{fig:polar}~(g) and (k) shows that the perpendicular component dominates the emitted harmonic intensity in this case. In Figs.~\ref{fig:polar}~(d), (h), and (l), the emitted fifth harmonic at the most efficient angle is dominated by the perpendicular component, in spite of the fact that the parallel component is dominant at all angles in the absence of a static field. Hence, the emission paths of the perpendicular components are expected to be important for the enhancement of MIR-induced HHG by a THz field. We observe the same trend for higher-order harmonics (see Appendix~\ref{sec:angle-dep-higher}).

\subsection{Comparison of the nonequilibrium steady state and the thermodynamic model}

Here, we investigate the role of a nonequilibrium steady state in HHG by comparing the results of the quasistatic approximation and the thermodynamic model~\cite{mics2015thermodynamic}. As introduced in Sec.~\ref{subsec:quasistatic}, the quasistatic approximation consists of replacing the THz pulse with the corresponding static field to describe the electronic system under a THz field. By contrast, the thermodynamic model consists of approximating the electronic system under a THz pulse by a state with a high electron temperature~\cite{mics2015thermodynamic}. The difference between the quasistatic approximation and the thermodynamic model reflects the difference between the nonequilibrium and equilibrium distributions, clarifying the role of the nonequilibrium steady-state in HHG.

The quasistatic approximation is characterized by the strength of the static field, $E_{\dc}$, whereas the thermodynamic model is characterized by the electron temperature $T_e$. To compare these two models that are formulated using different parameters, we introduce the excess energy~\cite{PhysRevB.106.024313} as a common measure of the excitation intensity. The excess energy corresponding to the quasistatic approximation is computed as the change in the total energy given in Eq.~(\ref{eqn:totalenergy}) of the nonequilibrium steady state caused by the application of the static field $\vecb E_{\dc}(t)$. Hence, the excess energy of the nonequilibrium steady-state depends on the static field strength as $\Delta E^{\mathrm{non-eq}}_{\textrm{excess}}(E_{\dc})$.  The excess energy of the thermodynamic model is computed as the change in the total energy caused by increasing the temperature from room temperature ($T_e=300$~K) and hence, depends on the electron temperature as $\Delta E^{\textrm{thermo}}_{\textrm{excess}}(T_e)$. Thus, converting both the static field strength $E_{\dc}$ in the quasistatic approximation and the electron temperature $T_e$ in the thermodynamic model to common excess energy enables the two models to be objectively and quantitatively compared~\cite{PhysRevB.106.024313}.

\begin{figure}[ht]
\includegraphics[width=0.9\linewidth]{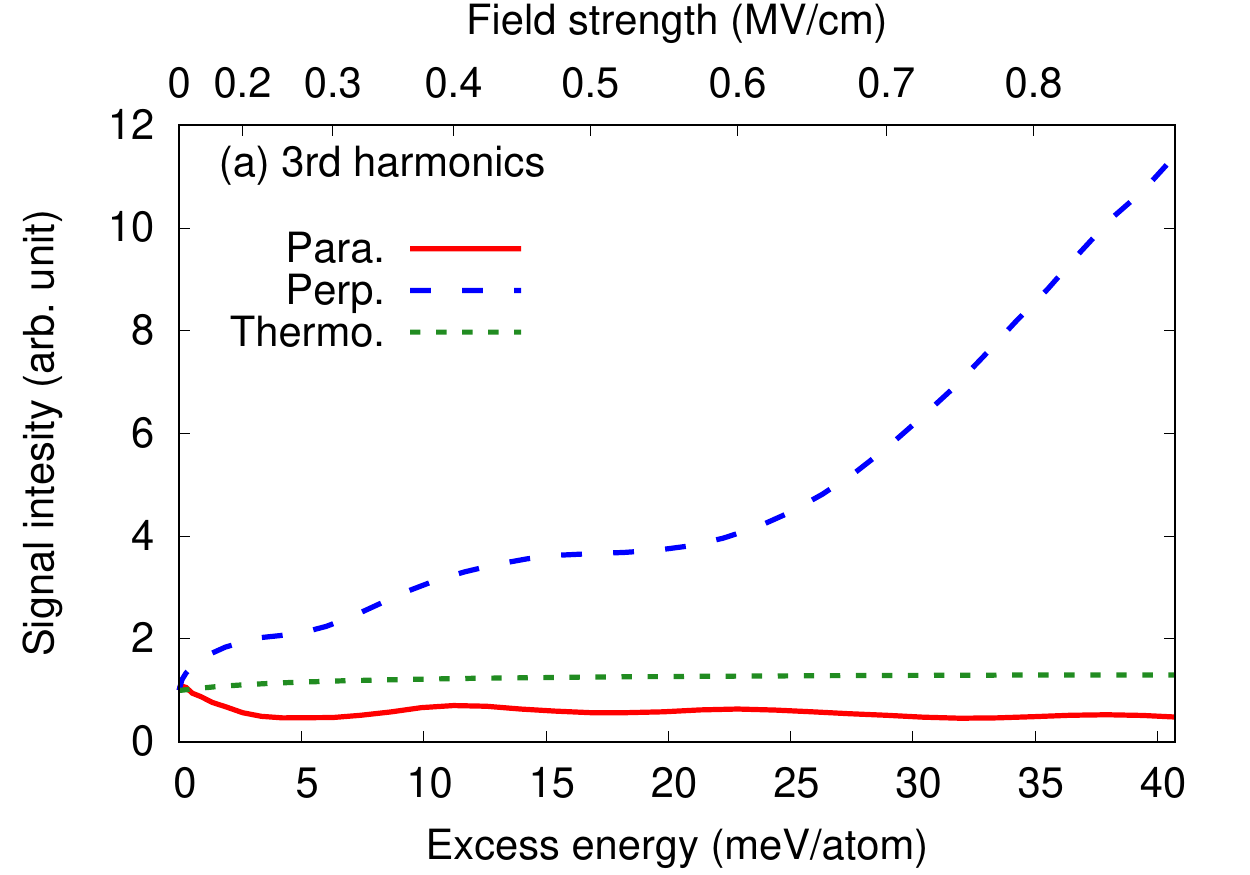}
\includegraphics[width=0.9\linewidth]{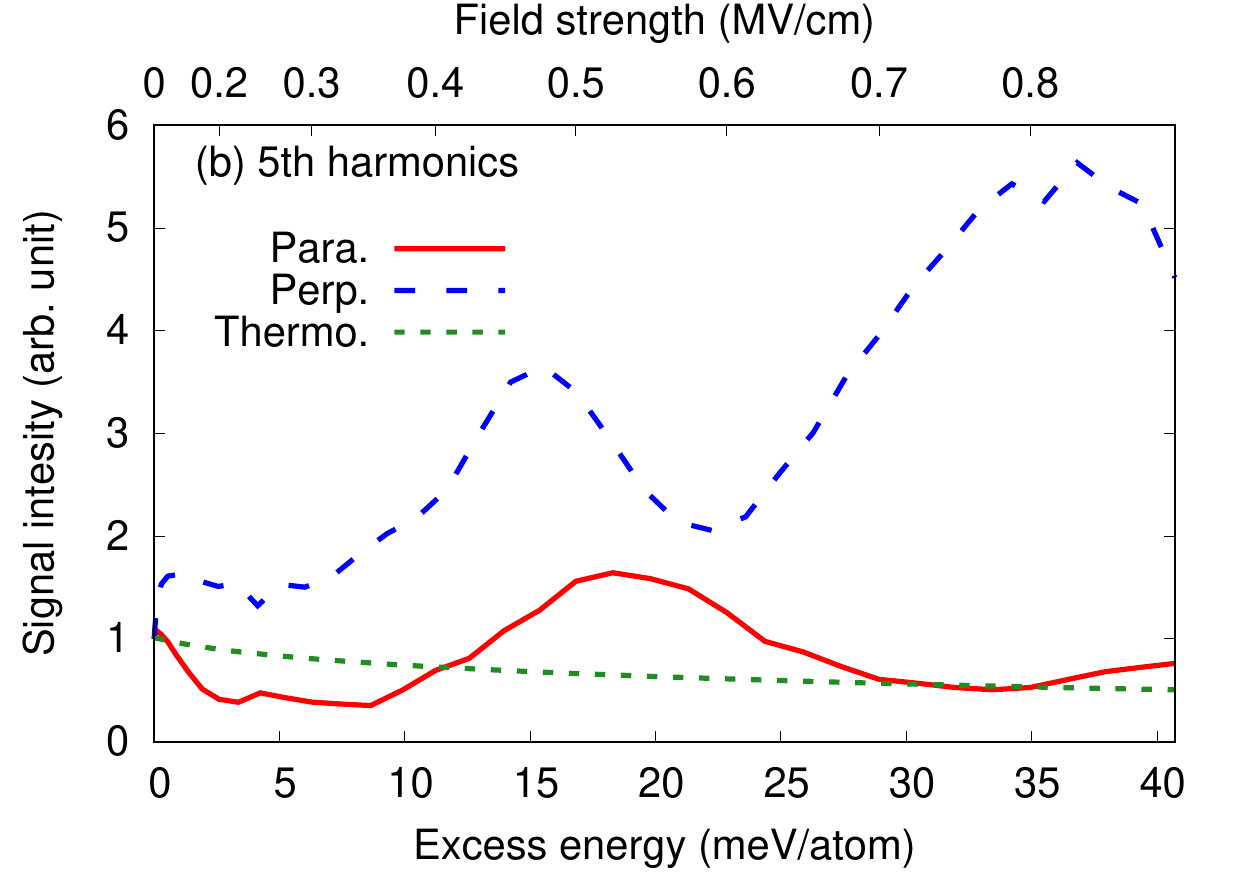}
\includegraphics[width=0.9\linewidth]{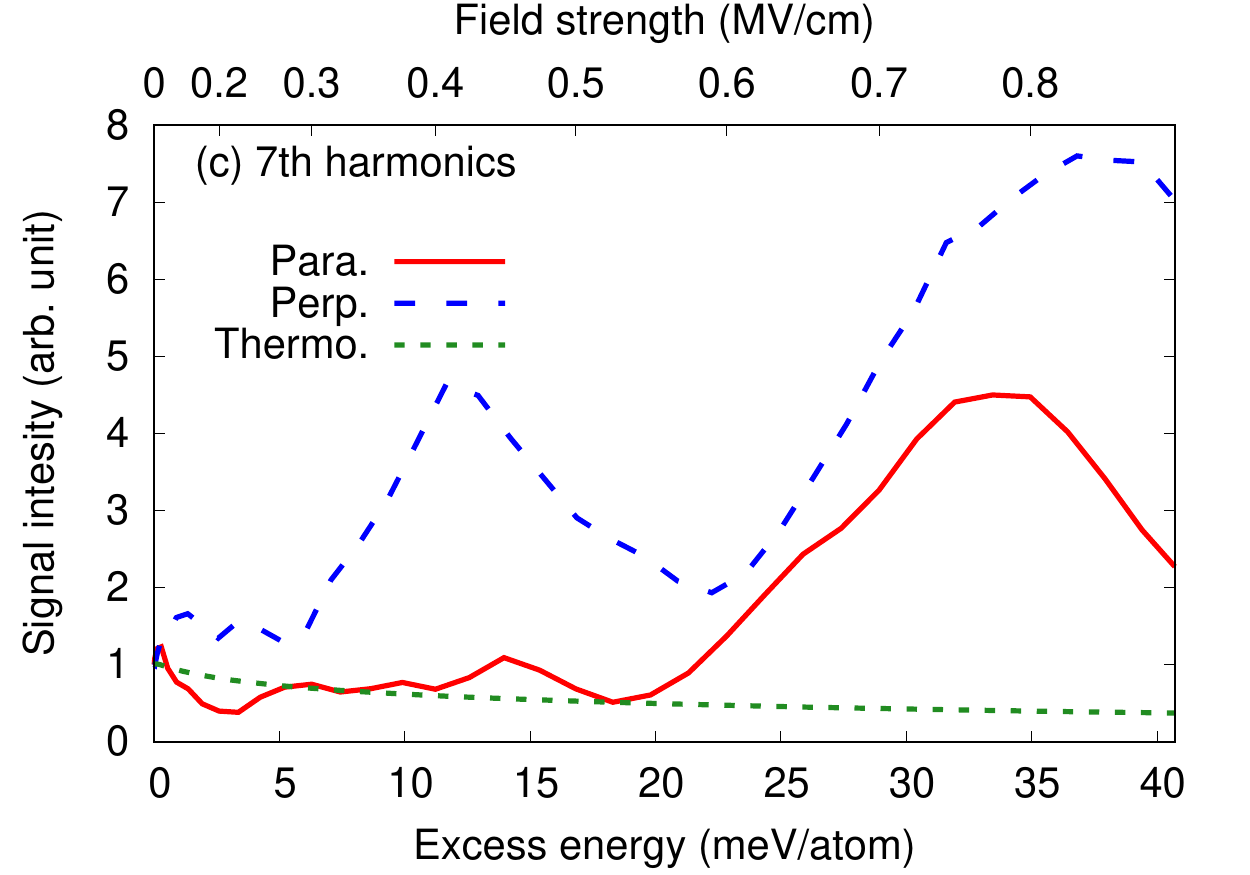}
\caption{\label{fig:intensity_tem}
The emitted light intensity, $I^{\textit{n}th}$, is shown as a function of the excess energy for (a) third (b) fifth, and (c) seventh harmonics. The results for the nonequilibrium steady-states induced by a static field parallel (red solid line) and perpendicular (blue dashed line) to the MIR field are compared with the thermodynamic model (green dotted line). In each panel, the field strength of the static field parallel to the MIR field is shown as the secondary axis.
}
\end{figure}

Figure~\ref{fig:intensity_tem} shows the comparison of the results obtained using quasistatic approximation and the thermodynamic model. The MIR field strength is set to 6.5~MV/cm, and the MIR field polarization direction is set to the $\Gamma$--$M$ direction (the $x$-axis of the present setup). The thermodynamic model preserves the intrinsic inversion symmetry of graphene and therefore does not produce even-order harmonics. Hence, we only analyze the odd-order harmonics generated by this model. Figure~\ref{fig:intensity_tem}~(a) shows that under a static field, perpendicular and parallel to the MIR polarization, the MIR-induced third harmonic is considerably enhanced and suppressed within the quasistatic approximation but remains almost constant within the thermodynamic model. Figures~\ref{fig:intensity_tem}~(b) and (c), show that under a static field, the fifth- and seventh-harmonic yields are significantly enhanced within the quasistatic approximation but decrease slightly as the electron temperature increases within the thermodynamic model. Hence, the enhancement of HHG cannot be described by simple heating of electronic systems within the thermodynamic model and originates from the non-equilibrium nature of field-induced electronic dynamics. The small change in the harmonic yields within the thermodynamic model relative to that predicted by the nonequilibrium steady state picture indicates that modification of the population distribution around the Fermi level has little effect on the spectra of HHG.

\subsection{Contribution of the nonequilibrium population}

Having demonstrated the importance of the nonequilibrium steady-state under a THz field, we elucidate the role of a \textit{coherent} coupling between the MIR and THz fields beyond the simple population contribution induced by the THz field. To highlight the coherent coupling contribution, we evaluate the contribution from \textit{incoherent} coupling by introducing a nonequilibrium population distribution model as an extension of the thermodynamic model. Within the thermodynamic model, the contribution from the THz field is described by modifying the population distribution by increasing the electronic temperature of the reference Fermi--Dirac distribution. Hence, the thermodynamic population only captures the population contribution (the diagonal element of the density matrix) of the THz-induced effect based on the thermal distribution. Here, we extend the thermodynamic model by replacing the reference Fermi--Dirac distribution in the relaxation operator in Eq.~(\ref{eqn:relaxation}) with the population distribution of the nonequilibrium steady state under a static field. The extended model includes the population contribution (given by diagonal elements of the density matrix) but not THz-induced coherence (given by the off-diagonal elements of the density matrix). Hence, a comparison of the nonequilibrium population model and the fully dynamical model can reveal the contribution from the coherent coupling between the THz and MIR fields.

To formulate the nonequilibrium population model, we first analyze the population distribution in the nonequilibrium steady state under a static field. The population distribution in the Brillouin zone can be expressed as
\begin{align}
n_{b \vecb k}(t) & = \int d \vecb k' \delta(\vecb k - \vecb K'(t)) \mathrm{Tr}\left [
| u^H_{b\vecb k'}(t)\rangle \langle u^H_{b\vecb k'}(t)| \rho_{\vecb k'}(t) 
\right ] \nonumber \\
&=\langle u^H_{b,\vecb k-e\vecb A(t)}(t)| \rho_{\vecb k-e\vecb A}(t) 
| u^H_{b,\vecb k-e\vecb A(t)}(t)\rangle,
\end{align}
where $\vecb K'(t)$ is the accelerated wavevector in accordance with the acceleration theorem, $\vecb K'(t)=\vecb k'+e\vecb A(t)$. The population distribution in the nonequilibrium steady state can be evaluated in the long-time propagation limit under a static field $\vecb A(t)= \vecb E_{\dc}\Theta(t)$,
\begin{align}
n^{\mathrm{neq-steady}}_{b\vecb k} = \lim _{t\rightarrow \infty } n_{b \vecb k}(t).
\end{align}

\begin{figure}[ht]
\includegraphics[width=0.9\linewidth]{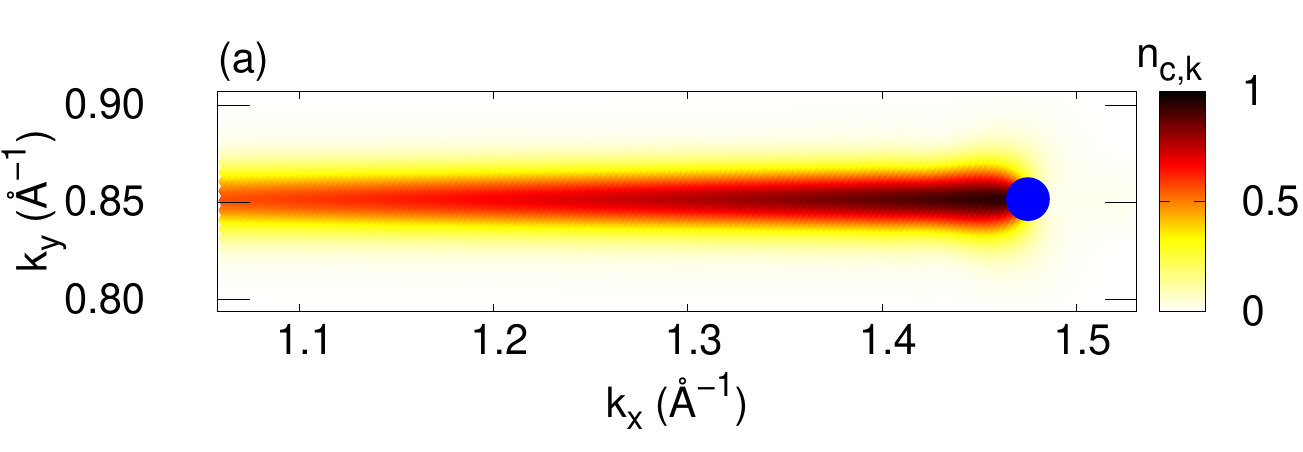}
\includegraphics[width=0.9\linewidth]{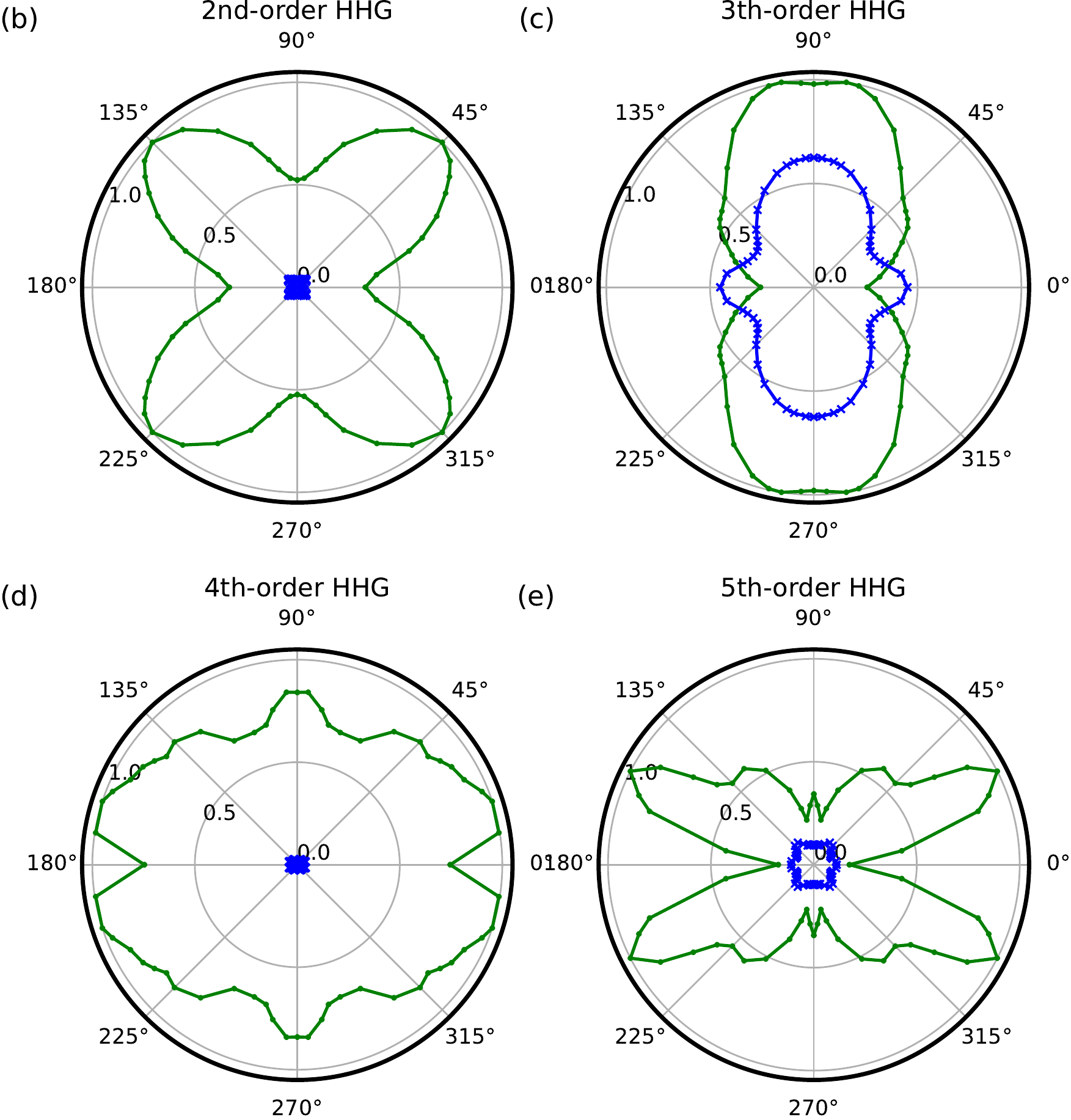}
\caption{\label{fig:pop}
(a) The calculated conduction population distribution, $n^{\mathrm{neq-steady}}_{c\vecb k}$ for the nonequilibrium steady-state is shown. Here, the Dirac point is indicated by the blue circle. (b--e) The angular dependence of the emitted harmonic intensity is shown for the (b) second, (c) third, (d) fourth, and (e) fifth harmonics. The results obtained using the nonequilibrium population model and the nonequilibrium steady-state are shown by the blue and green solid lines, respectively.
}
\end{figure}

Figure~\ref{fig:pop}~(a) shows the population distribution in the conduction band for the nonequilibrium steady-state under a static field with a strength of $E_{\dc}=0.5$~MV/cm. The static field is polarized along the $\Gamma$--$M$ direction ($x$-axis). The Dirac point ($K$ point) is depicted by the blue circle. In Fig.~\ref{fig:pop}~(a), the region to the left of the Dirac point is largely occupied by the field-induced population in the nonequilibrium steady-state, whereas the region to the right of the Dirac point is almost empty, breaking the inversion symmetry of the system. We employ this population distribution as the reference distribution of the relaxation operator in Eq.~(\ref{eqn:relaxation}) instead of the Fermi--Dirac distribution used to construct the nonequilibrium population model in the aforementioned discussion.

Figure~\ref{fig:pop}~(b) shows the angular dependence of the second-harmonic yield in the presence of a static field with a strength of $E_{\dc}=0.5$~MV/cm. The corresponding angular dependences of the third, fourth, and fifth harmonics are shown in Figs.~\ref{fig:pop}~(c--e), respectively. In each panel, the result obtained using the nonequilibrium population model is shown by the blue solid line, and the green solid line corresponds to the result obtained using the fully dynamical model, which is identical to the result shown in Fig.~\ref{fig:polar}. In Figs~\ref{fig:pop}~(b) and (d), the even-order harmonics computed with the nonequilibrium population model are negligibly weak compared with those computed using the fully dynamical calculation. This result indicates that under the charge-neutral condition ($\mu=0$) investigated here, the two- and four-phonon resonances of the MIR field are far from the Fermi level and cannot be modified by the population changes around the Fermi surface, resulting in small contributions to even-order harmonic generations. By contrast, within the fully dynamic calculation, the THz field can coherently couple with the MIR field via the off-diagonal elements of the density matrix. Thus, the coherent coupling can be realized both around the Fermi level and anywhere in the Brillouin zone, as long as the dipole transition is allowed. Hence, the coherent coupling contribution may enhance the contribution from the resonant quantum pass, inducing stronger even-order harmonic generation.

Figure~\ref{fig:pop}~(c) shows that for perpendicular THz and MIR fields, the third-harmonic yield calculated using the fully dynamical model is $1.57$ times stronger than that computed using the nonequilibrium population model when the two fields are perpendicular. This result indicates that the THz field enhances third-harmonic generation for the perpendicular configuration and that both coherent coupling and the incoherent population play important roles in the THz-assisted enhancement mechanism. By contrast, the third-harmonic yield computed using the fully dynamical calculation is $0.57$ times smaller than that computed using the nonequilibrium population model when the two fields are parallel. This result indicates that the contributions from the coherent coupling and the incoherent population cancel each other, weakening the total signal. Therefore, both coherent coupling and incoherent population affect third-harmonic generation under the investigated condition but play different roles depending on the relative angle $\theta$ between the THz and MIR fields.

Figure~\ref{fig:pop}~(e) shows that the fifth-order harmonic yield computed using the fully dynamical calculation is considerably larger than that computed using the nonequilibrium population model, except in the range where the THz and MIR fields are parallel. Hence, the coherent coupling is the dominant contribution to the enhancement of fifth-harmonic generation for most angles but the effects of coherent coupling and the incoherent population are both important when the MIR and THz fields are parallel. The consistent results are obtained for higher-order harmonics (see Appendix~\ref{sec:angle-dep-higher}).

\section{Summary \label{sec:summary}} 

We used a quantum master equation to model MIR-induced HHG in graphene in the presence of a strong THz field. We first computed the electron dynamics in graphene by explicitly employing MIR and THz pulses and evaluated the emitted high-harmonic spectra. Next, we developed a quasistatic approximation by analyzing MIR-induced HHG under a static field to replace the THz pulses. The THz-assisted MIR-induced HHG spectra were accurately reproduced by a static field within the quasistatic approximation, thus validating the application of this approximation for describing the induced dynamics generated by an applied strong THz field.

We then investigated the intensity of the emitted harmonics for different relative angles between the MIR and THz fields. In the absence of a THz field, the emitted odd-order harmonics exhibit an almost circular angular dependence, reflecting the circular symmetry of Dirac cones, whereas no even-order harmonics are emitted because of the intrinsic inversion symmetry of graphene. Under an intense THz field, the emitted harmonics exhibit a strong angular dependence along with enhancement and suppression of the harmonic yield. For example, the emitted fifth harmonic can be enhanced 10 times under a THz field with a strength of 0.5~MV/cm with respect to the result without the THz field, as shown in Fig.~\ref{fig:polar}~(d).

To elucidate the mechanism by which a THz field enhances MIR-induced HHG, we compared the results obtained using the quasistatic approximation and the thermodynamic model, which treats the effect of the THz field as a simple increase in the electron temperature of the Fermi--Dirac model~\cite{mics2015thermodynamic}. The thermodynamic model does not reproduce the enhancement of MIR-induced HHG, indicating that nonequilibrium THz-induced dynamics play an essential role in the enhancement.

To gain further insight into the enhancement of MIR-induced HHG by a THz field, we developed a nonequilibrium population distribution model. Within this model, THz-induced effects are treated as a change in the population distribution in the nonequilibrium steady state. The results obtained using the fully dynamical calculation and the nonequilibrium population distribution model were compared to elucidate the roles of coherent coupling between the MIR and THz fields. The THz-induced even-order harmonics and the THz-enhanced high-order harmonics are dominated by the coherent coupling contribution, whereas the enhancement of the third harmonics under a THz field is affected by both the coherent coupling and the nonequilibrium population. Furthermore, the enhancement of the higher-order harmonics is dominated by the coherent coupling contribution. These enhancement mechanisms are not rigidly limited by the conditions of the laser parameters investigated in this work but can be induced in rather general conditions. Therefore, it is key to control both the coherent coupling and the population for enhancing HHG from solids by employing multicolor laser fields.

\begin{acknowledgments}
The authors acknowledge fruitful discussions with K.~Nakagawa, H.~Hirori, and Y.~Kanemitsu.
This work was supported by JSPS KAKENHI Grant Numbers JP20K14382 and JP21H01842, the Cluster of Excellence 'CUI: Advanced Imaging of Matter'- EXC 2056 - project ID 390715994, SFB-925 "Light induced dynamics and control of correlated quantum systems" – project 170620586  of the Deutsche Forschungsgemeinschaft (DFG),  and the Max Planck-New York City Center for Non-Equilibrium Quantum Phenomena. The Flatiron Institute is a division of the Simons Foundation.
This work used computational resources of the HPC systems at the Max Planck Computing and Data Facility (MPCDF) and the Fujitsu PRIMERGY CX400M1/CX2550M5 (Oakbridge-CX) at the Information Technology Center, The University of Tokyo through the HPCI System Research Project (Project ID:hp220112).
\end{acknowledgments}

\appendix
\section{Angular dependence of high-order harmonics without THz fields \label{sec:angle-dep-higher-wo-thz}} 

\begin{figure}[ht]
\includegraphics[width=0.60\linewidth]{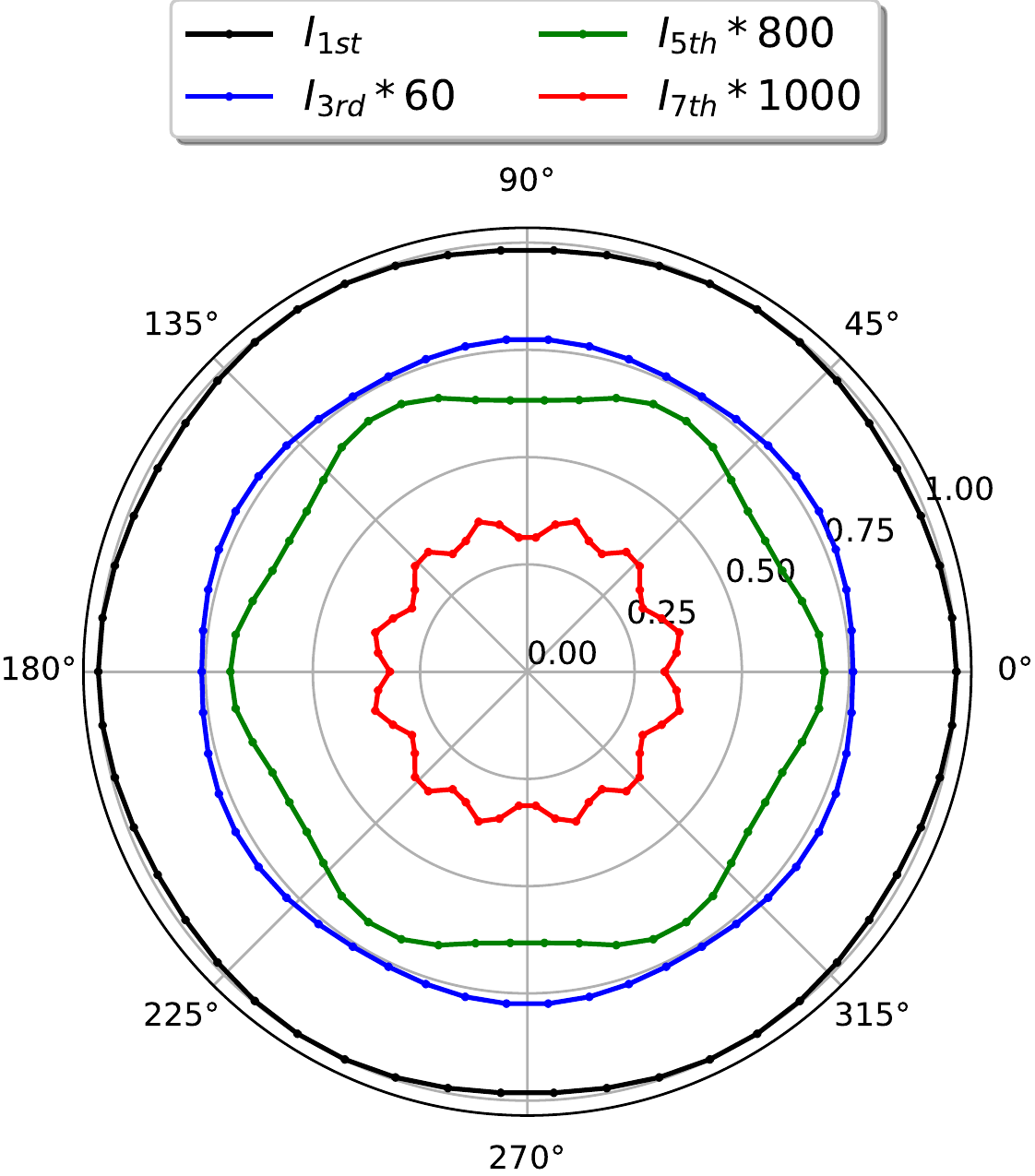}
\caption{\label{fig:SI_polar_mir}
The angular dependence of the harmonic yield obtained from the electron dynamics calculations in the presence of the MIR field. The numerical conditions are the same as those of the calculations in Fig.~\ref{fig:polar}. The third, fifth, and seventh harmonic yields are scaled by factors of 60, 800, and 1000, respectively.
}
\end{figure}

Here, we evaluate the angular dependence of the emitted harmonic yield $I^{n \textrm{th}}$ without THz fields, analyzing the intrinsic symmetry of graphene. Figures~\ref{fig:SI_polar_mir} show the computed angular dependence of the emitted harmonic yield $I^{n \textrm{th}}$ obtained using only the MIR field in the same conditions as Fig.~\ref{fig:polar}. Reflecting the six-fold symmetry of the hexagonal lattice of graphene, the emitted harmonics also show the six-fold symmetry in the angular dependence. As seen from Fig.~\ref{fig:SI_polar_mir}, the lower-order harmonics exhibit an almost circular angular dependence, reflecting the circular symmetry of Dirac cones. By contrast, the higher-order harmonics exhibit more complex six-fold symmetry in the angular dependence since the electronic structure of graphene deviates from a simple Dirac cone when a single-particle energy is far from the Dirac point.

\section{Angular dependence of high-order harmonics \label{sec:angle-dep-higher}} 

\begin{figure}[ht]
\includegraphics[width=0.95\linewidth]{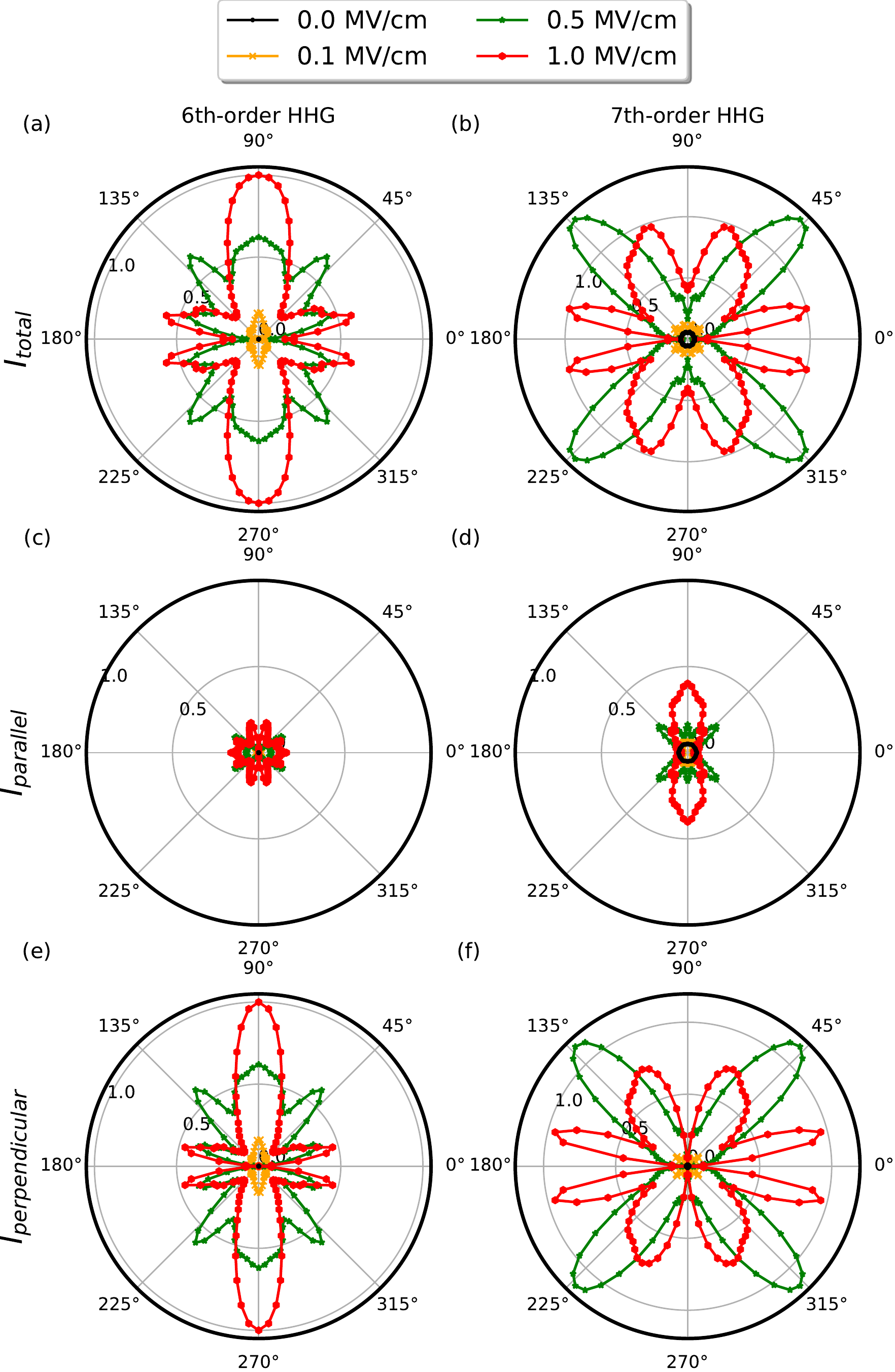}
\caption{\label{fig:SI_polar}
The angular dependence of the harmonic yields in the nonequilibrium steady state under a static field along the $\Gamma$--$M$ direction is shown. The angle $\theta$ denotes the relative angle between the static field and the $\MIR$ field. (a and b) The total intensity $I^{n \textrm{th}}_{\mathrm{total}}$ for the sixth and seventh harmonics is shown, respectively. (c and d) The component of the intensity parallel to $\vecb e_{\MIR}$ is shown for each harmonic. (e and h) Th component of the intensity perpendicular to $\vecb e_{\MIR}$ is shown for each harmonic. The results are normalized by the maximum total intensity $I^{n \textrm{th}}_{\mathrm{total}}$ of each harmonic.
}
\end{figure}

Here, we analyze the angular dependence of the high-order harmonics in the same way as that used to analyze Fig.~\ref{fig:polar}. Figures~\ref{fig:SI_polar}~(a) and (b) show the angular dependence of the sixth and seventh harmonics, respectively. Figures~\ref{fig:SI_polar}~(c) and (e) show the sixth-harmonic signal decomposed into parallel and perpendicular components, respectively. The same decomposition is shown for the seventh-order harmonic in Figs.~\ref{fig:SI_polar}~(d) and (f).

Consistent with the results for the fourth and fifth harmonics shown in Fig.~\ref{fig:polar}, the perpendicular components make a large contribution to the enhancement of MIR-induced HHG by a THz field, as shown in Fig.~\ref{fig:SI_polar}.

\begin{figure}[ht]
\includegraphics[width=0.95\linewidth]{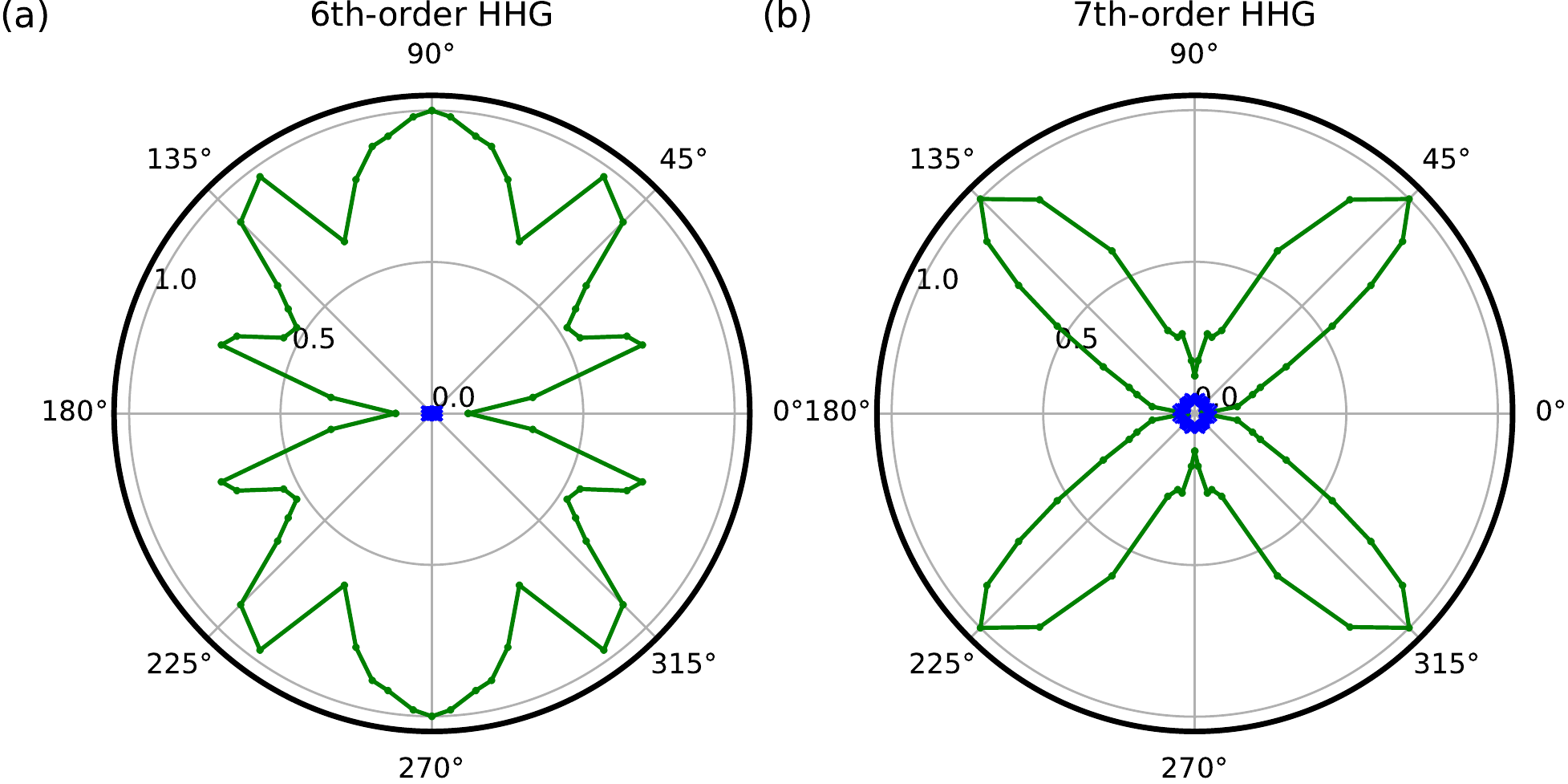}
\caption{\label{fig:SI_pop}
The angular dependence of the emitted harmonic intensity for the (a) sixth and (b) seventh harmonics are shown. The results obtained using the nonequilibrium population model and the nonequilibrium steady-state are shown by the blue and green solid lines, respectively.
}
\end{figure}

Furthermore, we compare the results for the sixth and seventh harmonics obtained using the nonequilibrium population model and the nonequilibrium steady state. Figures~\ref{fig:SI_pop}~(a) and (b) show the angular dependence of the sixth- and seventh-harmonic yields in the presence of a static field with a strength of $E_{\dc}=0.5$~MV/cm, respectively

Consistent with the analysis results shown in Fig.~\ref{fig:pop}, the coherent coupling between the MIR and THz fields plays an essential role in the enhancement of the HHG and goes beyond the simple field-induced population contribution.

\clearpage
\bibliography{ref}


\end{document}